\documentclass{emulateapj}

\usepackage{times}
\usepackage[]{graphicx,epsfig,rotating}

\shorttitle{A Fundamental Line for Elliptical Galaxies}
\shortauthors{Nair, van den Bergh, \& Abraham}

\begin{document}

\bibliographystyle{apj}

\title{A Fundamental Line for Elliptical Galaxies}

\author{
 Preethi Nair\altaffilmark{1,3}, Sidney van den Bergh\altaffilmark{2}, Roberto G. Abraham\altaffilmark{3}
}

\altaffiltext{1}{INAF - Astronomical Observatory of Bologna, 
                            Via Ranzani 1, I - 40127 Bologna, Italy;
                            preethi.nair@oabo.inaf.it}
                            
\altaffiltext{2}{Dominion Astrophysical Observatory,
Herzberg Institute of Astrophysics,
National Research Council of Canada,
5071 West Saanich Road,
Victoria, British Columbia, V9E 2E7
Canada;
sidney.vandenbergh@nrc-cnrc.gov.ca }

\altaffiltext{3}{
   Department of Astronomy \& Astrophysics,
   University of Toronto, 50 St. George Street,
   Toronto, ON, M5S~3H4, Canada
   abraham@astro.utoronto.ca
}

\slugcomment{Submitted to ApJ Letters on March 10, 2011; Accepted on May 10, 2011}


\begin{abstract}
Recent studies have shown that massive galaxies in the distant universe are surprisingly compact, with typical sizes about a factor of three smaller than equally massive galaxies in the nearby universe. It has been suggested that these massive galaxies grow into systems resembling nearby galaxies through a series of  minor mergers. In this model the size growth of galaxies is an inherently stochastic process, 
and the resulting size-luminosity relationship is expected to have considerable environmentally-dependent scatter.
To test whether minor mergers can explain the size growth in massive galaxies, we have closely examined the scatter in the size-luminosity relation of nearby elliptical galaxies using a large new database \citep{2010ApJS..186..427N}  of accurate visual galaxy classifications. 
We demonstrate that
this scatter is much smaller than has been previously assumed,
and may even be so small as to challenge the plausibility of the merger driven
hierarchical models for the formation of massive ellipticals.
 \end{abstract}

\section{INTRODUCTION}
\label{sec:introduction}


Galaxies exhibit a bewildering variety of shapes and sizes, but
elliptical galaxies are the simplest of all extragalactic objects, 
and their structural uniformity suggests that they might be 
easier to understand than more complex later type galaxies.
Most recent attempts to understand the evolution of elliptical galaxies
have been in the context of hierarchical models for structure
formation \citep{Toomre:1977p28807,White:1978p17918,Khochfar:2003p28813,Bower:2006p23715,DeLucia:2006p23815,Hopkins:2010p28867}. In these models the formation of elliptical galaxies is linked
to mergers, where the galaxy's environment is the central parameter which determines its merger history \citep{Mo:1996p22126}. 
The predominance of old stellar populations in nearby massive elliptical galaxies 
is inconsistent with large episodes of recent star formation, which
suggests that the mergers which formed elliptical galaxies either
occurred very long ago (at redshifts $z>$2), or else that they can best be described
as the coalescence of pre-existing old stellar populations (`dry mergers') \citep{Thomas:2005p22786,Renzini:2006p19035,Graves:2009p22588,Graves:2009p22581}. 
Such dry mergers do not form many new stars; instead they simply rearrange the existing stars. 
It has been argued that such dry merging 
may also be responsible for the observed large size growth in
massive, compact, elliptical galaxies  over the last 10 Gyrs  \citep{Cimatti2004, Daddi2005, Longhetti2007, vanDokkum:2008p2119, McGrath2008,2009ApJ...695..101D, Rettura2010, 2009ApJ...691.1424H, 2009ApJ...699L.178N, 2008ApJ...688...48V}, 
though this scenario requires considerable fine tuning to 
reproduce local scaling relations
\citep{2009ApJ...706L..86N}.  
This growth is also expected to be dependent on environment, with galaxies
in higher density environments undergoing more mergers and hence showing a larger scatter in their size-luminosity relation.

Recent studies \citep{1996A&A...312..397G,2008A&A...490...97V,2010ApJ...715..606N} 
have found that the size-luminosity relation provides the tightest of all the purely photometric correlations 
used to characterize galaxies. 
However, the environmental dependence of this relation is highly debated with some groups finding a large scatter, 
as well as a strong environment dependent curvature in the size-luminosity relation for elliptical galaxies \citep{Giuricin:1988p12909,Shen:2003p70,2007AJ....133.1741B,2007MNRAS.377..402D,2007MNRAS.379..867V,2009MNRAS.394.1978H},
while others do not \citep{Guo:2009p19543, Weinmann:2009p19068, 2010ApJ...715..606N}.
The purpose of the present paper is to show that the size-luminosity relation of elliptical galaxies
is well defined by a fundamental line with no environmental dependence. 
We demonstrate that
this scatter is much smaller than has been previously assumed,
and may even be so small as to challenge the plausibility of the merger driven
hierarchical models for the formation of massive ellipticals.

Throughout our paper we assume a flat dark energy-dominated cosmology with $h=0.7$, $\Omega_m=0.3$ and $\Omega_\Lambda$=0.7. 


\section{SAMPLE}

\subsection{Main Sample}
Our sample of 2861 elliptical galaxies is a subset of the  $14,034$ visually classified bright galaxies (model $g^{\prime}<16$, $0.01<z<0.1$) presented in \cite{2010ApJS..186..427N} (NA10) with valid measures of environment (described below) and no overlapping companions. The NA10 sample is in turn derived from the Sloan Digital Sky Survey spectroscopic catalog \citep{2002AJ....124.1810S}. Visual classification was carried out by one of the authors (P.N.) and found to agree with those from the RC3 within one Hubble type (on average) for the 1,793 galaxies in common to both samples. 
It is important to emphasize that, unlike samples derived from automated classifications, our sample is expected to contain only minimal contamination by S0 and Sa galaxies. 
For this analysis, we updated the photometry and spectroscopy to use the SDSS Data Release 7 (DR7) derived sizes, luminosities and velocity dispersions.

The sizes of galaxies are parameterized by (a) the radius enclosing 90\% of the galaxy's light contained within twice the Petrosian radius \citep{Petrosian:1976p6531,Stoughton:2002p1611}, referred to as $R_{90}$; (b) the corresponding radius containing 50\% of the galaxy's light, referred to as $R_{50}$; and (c) the seeing-corrected de Vaucouleurs  radius, also known as the half light radius, $R_{e}$. The de Vaucouleurs radii are corrected for the known problem originating in 
slight errors in SDSS sky-subtraction of large galaxies, as specified by \cite{2009MNRAS.394.1978H}~~[HB]. 

We explored two estimators of galaxy luminosity.
The total magnitude of each galaxy was specified by  (a) the Petrosian luminosity and (b) the de Vaucouleurs luminosity, both corrected for extinction,  k-correction \citep{2005AJ....129.2562B} and luminosity evolution \citep{2009MNRAS.394.1978H}, and the latter also corrected for the known problems of SDSS sky-subtraction as prescribed by \cite{2009MNRAS.394.1978H}. All magnitudes in the present paper are in the AB system.

It is important to understand the size of the intrinsic scatter in our measurements.
As will be shown, the tightest relations are obtained using $R_{90}$.
From simulations, we estimate the fractional uncertainties in $R_{90}$ to be several times the fractional uncertainty in the sky level. On SDSS images the sky background around galaxies is generally known to about $\pm$ 1\%, so the corresponding fractional uncertainties in $R_{90}$ estimates are typically 4-5\%. Other error terms are added in quadrature to this, so that a typical $R_{90}$ measurement carries around a 5\% uncertainty. 
The effect of seeing correction on $R_{90}$ is negligible \citep[see Appendix A in][]{2010ApJ...715..606N}.
In comparison to the errors on the size measurements, the errors on both magnitude estimators are small ($<0.01$ mag)\footnote{ We ignore the systematic offset in magnitude caused by SDSS under-estimating sky for bright galaxies (r$<$14mag). However, if we apply the HB correction formula (which is an overestimate) directly to the Petrosian luminosity, the trends are similar.) } .  

Finally, we note that in the plots presented below we isolate subsets of galaxies based on central velocity dispersion. These velocity dispersions have been corrected for aperture effects as specified by \cite{Jorgensen:1995p23360}.  


\subsection{Environment Measures}
To study trends with environment, we use two publicly available metrics, 
an $N$th nearest neighbor approach computed by \cite{2006MNRAS.373..469B}, and a group catalog algorithm by  \cite{2007ApJ...671..153Y}. \cite{2006MNRAS.373..469B} measured the environmental density for  SDSS galaxies with 
$r^{\prime}<18$,  $0.01<z<0.1$
and photometrically selected galaxies with surface brightness 18.5$<$$\mu_{r,50}$$<$24.0. The density is defined as  $\Sigma = N/ (\pi d_{N}^{2})$, where $d_{N}$ is the projected comoving distance (in Mpc) to the Nth nearest neighbor. 
A best estimate density (to account for spectroscopic incompleteness) was obtained by calculating the average density for $N$=4 and $N$=5 with spectroscopically confirmed members only and with the entire sample. The mean log $\Sigma$ for our sample is -0.32.

\cite{2007ApJ...671..153Y} used an iterative halo-based group finder on the SDSS NYU-Value Added Galaxy Catalog \citep{2005AJ....129.2562B} for objects with 
$r^{\prime}<18$,
and 0.01$<$$z$$<$0.2 with a redshift confidence $C_{z}$$>$0.7. Tentative group members were identified using a modified friends-of-friends algorithm. The group members were used to determine the group center, size, mass, and velocity dispersion. New group memberships were determined iteratively based on the halo properties. The final catalog yields additional information identifying the brightest galaxy in the group (BCG), the most massive galaxy in the group (both used as proxies for central galaxies), estimated group mass, group luminosity, and halo mass.
We use the group occupation number $N$ as a proxy for environment. We are primarily interested in relative evolution between field and cluster galaxies and hence define low-density regions as galaxies with $N$$\le$2 and Baldry log $\Sigma$  $<$ -0.32, while high-density regions are defined as galaxies in groups with more than two members $N$$>$2 and Baldry log $\Sigma$  $>$ -0.32 (the mean $N$ and log $\Sigma$ for our sample). 

\section{SIZE-LUMINOSITY RELATION OF ELLIPTICAL GALAXIES}

\begin{figure*}
\begin{center}
\includegraphics[width=6.3in]{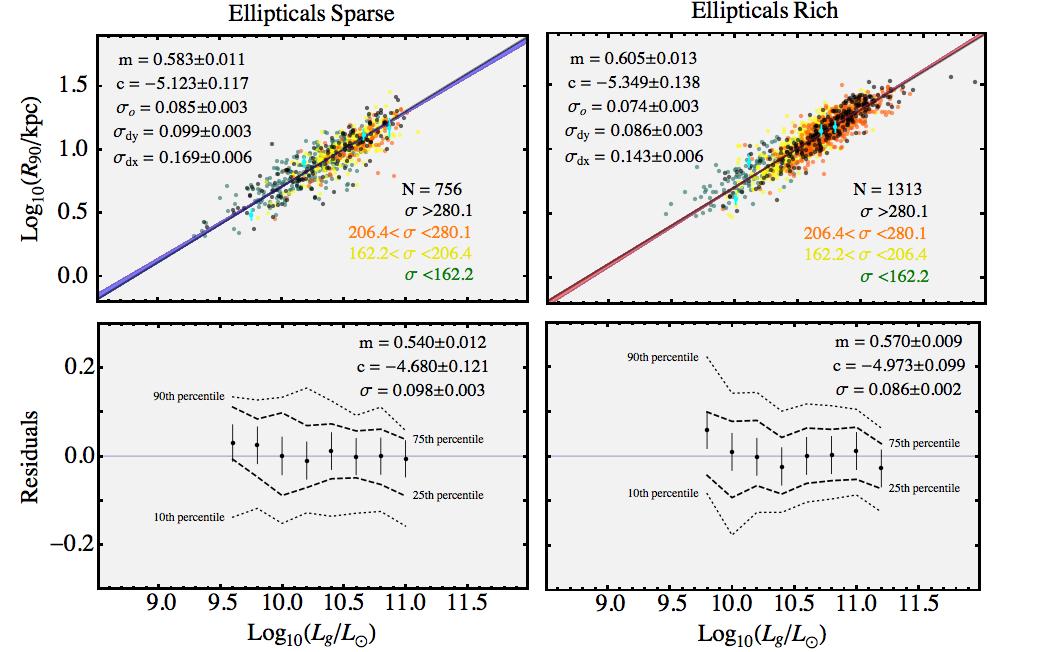}
\caption{
\sffamily\noindent
\sffamily\textbf{Petrosian Size-Luminosity relationships for nearby elliptical galaxies.}
The top row shows the relationships obtained from the \cite{2010ApJS..186..427N} elliptical galaxy
sample where the size of each galaxy is parameterized by the radius enclosing 90\% of the galaxy's light contained within
twice the Petrosian radius, $R_{90}$ (see text for details). The left-hand panel corresponds to galaxies in the field, while the right
hand panels correspond to galaxies in dense environments. 
Symbol colours are keyed to corrected central velocity dispersion of the galaxies, in four broad bins. In cyan are error bars for four random points. In each panel the best-fit orthogonal linear model (blue/red lines) is superposed on the data, with the parameters for the model inset. $\sigma_{dx}$ and $\sigma_{dy}$ are the scatter in the x and y parameters. The black line shows the best-fit relation for all elliptical galaxies.  
The bottom-row shows the residuals in size from the direct best fit analysis, with the parameters for the model inset. The dashed lines show the 25th percentile- 75th percentile range while the dotted lines show the  10th percentile- 90th percentile range. The error bars show the median error in size, including 1\% error in sky, in small bins of luminosity.
}
\label{default}
\end{center}
\end{figure*}

We determined the best-fitting size-luminosity relations for our sample by (a) minimizing the scatter in size (direct method) 
and (b) minimizing the scatter orthogonal to the best fit line (orthogonal method). 
While there is a slight change in fit parameters between the two methods, our overall conclusions are not affected.
In all cases we found that Petrosian-based quantities yielded considerably tighter relations than those
obtained with the `traditional' relations defined using de Vaucouluers profile fits.

Figure 1 (top panel) shows the Petrosian $R_{90}$ size- Petrosian luminosity relation for the 756 elliptical
galaxies that are located in sparse environments (left panel),
and the 1315 ellipticals in dense environments (right panel). 
The points are keyed to central velocity dispersion quartiles where black points are the galaxies with the highest velocity dispersion in each panel and green points have the lowest velocity dispersion. Orange and yellow points are the intermediate quartiles. The black line shows the best fit relation for all ($\sim$3000) elliptical galaxies.  The blue lines/regions and red lines/regions in each panel denote the best fit (orthogonal) relation in sparse and dense environments respectively, with the shaded region denoting the $1\sigma$ uncertainties in the slope (determined by 100 bootstraps). The parameters of the best fit orthogonal relation are inset in each panel and are summarized in Table~1.
The lower panels of Figure 1 show the  
residuals in size about the (direct) size-luminosity relation in small bins of luminosity. 
The parameters of the best fit direct relation are inset in each panel.
The error bars show the median error in size including a 1\% error in sky. 
The dashed contours indicate the 25th to 75th percentile range while the dotted lines indicate the range spanned by the 10th to the 90th percentiles.
The figure shows the following striking features: (1)
The radii and the luminosities of elliptical galaxies exhibit a very tight
power-law relationship (linear in log-log space) over a range of $\sim100$ in luminosity. 
(2) Within the 
statistical errors the elliptical galaxies in dense and in sparse environments appear 
to follow {\em the same} power-law relationship, 
though the scatter in this relation is slightly smaller in higher density environments. 
(3) The intrinsic scatter in the relationship
is comparable to the measurement errors in $R_{90}$\footnote{The measurement errors in $R_{90}$ are expected to be larger in higher density environments where SDSS suffers from over-estimation in sky background levels. This error is not included in the estimated errors.}.
Thus {\em the size-luminosity diagram of elliptical galaxies
defines a\, `fundamental line'}\, in log-log space with negligible intrinsic
scatter over two orders of magnitude in galaxy luminosity.

It is interesting to compare our results with those obtained by 
 \cite{2009MNRAS.394.1978H}~~[HB], who found curvature in the size-luminosity 
relationship.
The HB sample is
a factor of twenty larger than ours, but the sample was
chosen on the basis of automated classifications and 
is contaminated by
S0 and Sa galaxies. 
The authors analyzed their sample using conventional
galaxy sizes parameterized by the half-light radius $R_{e}$. Figure~2 presents the size-luminosity relationship of our 
sample analyzed using the Petrosian half-light radius ($R_{50}$) vs. Petrosian luminosity (top row) and $R_{e}$ vs. de Vaucouleurs luminosity (bottom row) keyed to central velocity dispersion quartiles. The color coding is the same as in Figure~1. The dashed black line shows the best fit relation from Figure~1.
A number of features
are apparent: (1) The slope, scatter and environmental dependence of elliptical galaxies using $R_{50}$ is nearly the same as $R_{90}$. 
For elliptical galaxies $R_{50}$ is just a factor of three smaller than $R_{90}$. Thus the ratio $R_{90}$/$R_{50}$ is not sensitive to the internal structure (S\'{e}rsic index) of elliptical galaxies.
(2) The curvature noted by HB (and clearly seen
in the lower panels) is not seen when using $R_{90}$ or $R_{50}$\footnote{This is also true when using the complete HB sample which probes higher luminosities.}.
This is because $R_{e}$, the half light radius, is sensitive to the profile shape (S\'{e}rsic index) of galaxies, as has been shown in Appendix A of \cite{2010ApJ...715..606N}. 
(3) The scatter in the size-luminosity relation using
$R_{90}$ (or $R_{50}$)  is $\sim40\%$ lower than that obtained with $R_{e}$.  

\begin{figure*}[htb!]
\begin{center}
\includegraphics[width=6.3in]{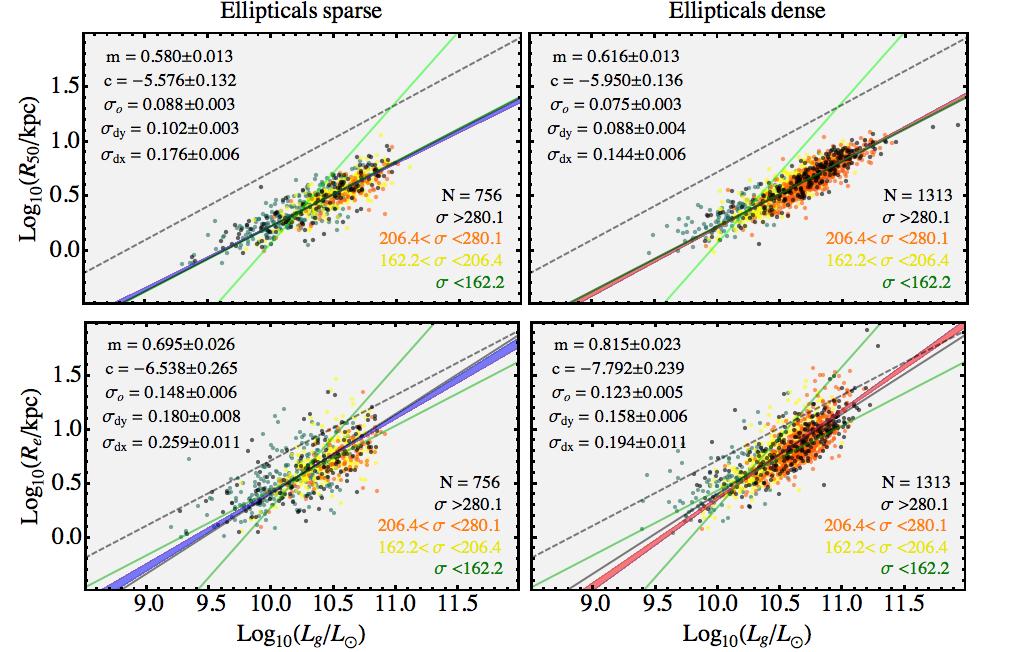}
\caption{
\sffamily\noindent
\sffamily\textbf{Decrease in scatter and curvature of the Size-Luminosity Relation obtained by using Petrosian $R_{90}$ ($R_{50}$) sizes instead of conventional sizes.}
The top row shows the relationships obtained using the
radii enclosing 50\% of the Petrosian flux, $R_{50}$
vs. Petrosian luminosity while the bottom-row shows the relationships obtained using de Vaucouleurs half-light radii, $R_{e}$,
vs. de Vaucouleurs luminosity.
The left-hand panels correspond to galaxies in the field, while the right
hand panels correspond to galaxies in dense environments, such as groups. 
Symbol colors are keyed to corrected central velocity dispersions of the galaxies, 
in four broad bins.  In each panel the best-fit orthogonal linear model is superposed
on the data, with the parameters for the model inset. $\sigma_{dx}$ and $\sigma_{dy}$ are the direct scatter in the x and y parameters. 
The solid black line shows the best-fit relation for all elliptical galaxies.  The dashed black line shows the best-fit relation from Figure~1.
The green lines indicate the predicted slopes from dry major merging assuming various orbital configurations from \cite{2006MNRAS.369.1081B}. The steeper line is for bound radial orbits, while the shallower line is for the bound orbit with the largest pericentric distance \citep{2006MNRAS.369.1081B}.}
\label{default}
\end{center}
\end{figure*}

\section{DISCUSSION}

The size-luminosity relation of elliptical galaxies is closely related to the {\em Fundamental Plane} 
\citep{1987ApJ...313...59D,1987nngp.proc..175F,1989ARA&A..27..235K},
which describes a rather tight
relationship between the size, surface brightness and central velocity dispersion
 of elliptical galaxies. The `tilt'  (with respect to canonical relationships predicted by the virial theorem)  
 and  scatter about the Fundamental Plane
are thought to be due to variations in either age, metallicity or structural non-homology.  
It is  remarkable that the scatter in the  Petrosian $R_{90}$ size-luminosity 
relationship ($\sim$0.090 dex in size for all elliptical galaxies) is 
tighter than the scatter recently reported 
in the Fundamental Plane ($\sim$0.097 dex in size $R_{e}$\,\citep{2009MNRAS.396.1171H}). 
How is it possible that the size, luminosity and velocity dispersion
information embedded in the fundamental {\em plane} formulation does not
provide a more accurate fit to the observations than does the fundamental {\em line}
defined by the size and luminosity data alone? 
Recent work on nearby Virgo cluster galaxies \citep{2009ApJS..182..216K}  
makes a strong case for the conclusion that the {\em internal structure} of
early-type galaxies depends on their minor merger history. 
If true, one might perhaps expect the dispersion about the fundamental plane to depend on the galaxy radius that is
chosen to define the plane. Figure~2 suggests that at least some of the tilt, scatter and curvature in the conventional fundamental plane 
may be due to structural (or kinematic) non-homology in the elliptical galaxy population with $R_{e}$,
and that this may disappear if the fundamental plane is defined using a metric of size which is not as sensitive to the galaxy profile shape. 
Note that this suggests that the record of the
galaxy's merger history may be most strongly imprinted in a galaxy's
profile, rather than in its overall size, which is what is being
probed by our investigation. 

It is interesting to consider if the small scatter in the size-luminosity relationship for
elliptical galaxies poses a challenge for theory. 
The merger theory of elliptical galaxies needs to account for both the tightness of the fundamental plane 
(and the tighter size-luminosity relation) and the growth of compact elliptical galaxies which account for up to 
$\thicksim$50\% of the galaxy population at redshifts of 2-3 \citep{vanDokkum:2008p2119}.
We compare our results to predictions from published simulations of
dry mergers \citep{2006MNRAS.369.1081B}. Figure~2 shows the predicted size-luminosity relationships 
 in green for 1:1 mass dry mergers
with various orbits  which preserve a fundamental plane (although not necessarily with the observed slope). The size calculated by \cite{2006MNRAS.369.1081B} is not a profile radius but the circular aperture enclosing half of the projected stellar mass, i.e. it is more comparable to $R_{50}$.
The steepest relation (slope = 1.2) is for the most radial orbit
while the shallower relation (slope = 0.6) has the largest pericentric separation between the merging pair. 
While the curvature in $R_{e}$ can be explained by radial dry merging,
the lack of curvature in $R_{90}$ and $R_{50}$ further suggests that $R_{e}$ is more influenced by profile changes induced by mergers.
The observed scatter (0.09 dex, see Table 1) in the $R_{90}$ ($R_{50}$)-luminosity relationship  seems much too small to be consistent with the predictions of dry merger
simulations. In fact, the observed scatter is consistent with the scatter due solely to orientation effects predicted by simulations \citep[0.1 dex with $10^{4}$ viewing angles, see][]{2006MNRAS.369.1081B}.
(Note that the scatter in luminosity is negligible assuming the galaxies are optically thin).
More worryingly, dry minor mergers are expected to introduce a similar and possibly larger scatter in size as do dry major mergers for a similar increase in mass (or luminosity) \citep{2009ApJ...706L..86N}. Variable gas fractions of the progenitors can introduce a further scatter in size \citep{2006ApJ...645..986R}.
It thus seems highly unlikely that major or minor mergers (either dry or wet) could be growing 
elliptical galaxies while preserving the slope and small scatter observed in the size-luminosity relation in both low and high density environments.

The current prevailing theory of size growth of distant galaxies
 can be described as the development of
an outer envelope which grows about a central dense `red nugget' \citep{2009ApJ...695..101D},
whose stellar density remains unchanged as the galaxy inflates \citep{2009ApJ...691.1424H,2009ApJ...697.1290B,2009ApJ...695..101D}.
The most straightforward expectation
\citep{2009ApJ...697.1290B}
 based on the virial theorem suggests that a minor merger
will grow the size of a galaxy in direct proportion to the additional mass added to
the total system by the galaxy being absorbed. There will be
considerable scatter in the growth depending on the relative orbital configurations
of the galaxies \citep{2006MNRAS.369.1081B}.
 If minor mergers are driving this process, not only must
the relative masses and orbital configurations of the merging systems be fine tuned in order to
grow galaxies while maintaining the slope of the size-luminosity relationship at all 
redshifts \citep{2009ApJ...706L..86N}, but the outcome must also preserve
a negligible scatter about the fundamental line over two orders of magnitude
in galaxy luminosity. Assuming a scatter of 0.07 dex in the size-mass relation, 
\cite{2009ApJ...706L..86N} find their simulations rule out
size growth larger than a factor of 1.9 by dry mergers.

In summary, it is concluded that the elliptical galaxy size-luminosity scaling relation determined using Petrosian $R_{90}$(or $R_{50}$ or $R_{p}$) has a much
smaller scatter than the same relation determined using the half-light radius $R_e$.
The relation
between the  Petrosian luminosities and Petrosian-based radii of elliptical galaxies is a simple
power law: $R_{90} \propto L^{0.6}$. The dispersion about this `fundamental line'  in log-log space is
found to be only 0.09 dex in size (0.36 mag in luminosity) in the local universe, smaller than that of the fundamental plane
defined using the effective radius $R_{e}$. 
The `fundamental line'  appears to be driven mainly by luminosity (or mass) and seems to be independent of environment 
with no curvature at higher luminosities. 
The observation that the structural properties of elliptical galaxies 
is both simple and independent of environment suggests that the theory of hierarchical
growth of elliptical galaxies via mergers is not understood. 
Is our paradigm for galaxy formation merely cracked or is it broken?

\acknowledgments
\noindent{\em Acknowledgments}
The authors thank the anonymous referee for comments and useful suggestions.
PN thanks Carlo Nipoti and Giovanni Zamorani for helpful discussions. 
Funding for SDSS has been provided by the Alfred P. Sloan Foundation, the Participating Institutions, NSF, the U.S. Department of Energy, NASA, the Japanese Monbukagakusho, the Max Planck Society, and the Higher Education Funding Council for England. The SDSS website is http://www.sdss.org/.


\begin{deluxetable}{rcccccc} 
\tablecolumns{7} 
\tablewidth{0pc} 
\tabletypesize{\scriptsize} 
\tablecaption{Size-Luminosity- Relationships\label{tab:relations}}
\tablehead{ 
 \colhead{} & \colhead{} &  \colhead{} & \colhead{} & \colhead{} &
  \multicolumn{2}{c}{-- Orthogonal Fit --        } \\  
  \colhead{Class} & 
  \colhead{$N$} &
  \colhead{Slope} &
  \colhead{Intercept} &
  \colhead{Dispersion} &
  \colhead{Dispersion R} &
  \colhead{Dispersion L \vspace{0.1cm}} \\
  \colhead{} & 
  \colhead{} &
  \colhead{$[\log(L_\odot)]$} &
   \colhead{$\left[\log(L_\odot) \over \log({\rm kpc)}\right]$ \vspace{0.1cm}} &
   \colhead{$[\log(L_\odot)]$} &
  \colhead{$[\log({\rm kpc})]$} &
   \colhead{$[\log({\rm kpc})]$}
} 
\startdata



 \cutinhead{\em Elliptical galaxies  $R_{90}$ vs. $L_{petrosian}$}    
     All & 2861 & $0.596 \pm 0.006$ & $-5.256\pm 0.062$ & $0.077 \pm 0.001$ & $0.090 \pm 0.002$ & $0.151 \pm 0.003$ \\   
  Sparse & 756 & $0.583 \pm 0.011$ & $-5.123 \pm 0.117$ & $0.085 \pm 0.003$ & $0.099 \pm 0.003$ & $0.169 \pm 0.006$ \\   
     Rich & 1313 & $0.605 \pm 0.013$ & $-5.349 \pm 0.138$ & $0.074 \pm 0.003$ & $0.086 \pm 0.003$ & $0.143 \pm 0.006$ \\



\cutinhead{\em Elliptical galaxies $R_{50}$ vs. $L_{petrosian}$}    
    All & 2861 & $0.596 \pm 0.006$ & $-5.746\pm 0.062$ & $0.080 \pm 0.001$ & $0.093 \pm 0.002$ & $0.156 \pm 0.003$ \\   
     Sparse & 756 & $0.580 \pm 0.013$ & $-5.576 \pm 0.132$ & $0.088 \pm 0.003$ & $0.102 \pm 0.003$ & $0.176 \pm 0.006$ \\   
     Rich & 1313 & $0.616 \pm 0.013$ & $-5.950 \pm 0.136$ & $0.075 \pm 0.003$ & $0.088 \pm 0.004$ & $0.144 \pm 0.006$ \\   

\cutinhead{\em Elliptical galaxies $R_{p}$ vs. $L_{petrosian}$}    
    All & 2861 & $0.6114 \pm 0.007$ & $-5.515\pm 0.070$ & $0.088 \pm 0.001$ & $0.103 \pm 0.002$ & $0.169 \pm 0.004$ \\   
     Sparse & 756 & $0.598 \pm 0.014$ & $-5.376 \pm 0.146$ & $0.097 \pm 0.003$ & $0.113 \pm 0.004$ & $0.189 \pm 0.007$ \\   
     Rich & 1313 & $0.630 \pm 0.014$ & $-5.711 \pm 0.149$ & $0.083 \pm 0.003$ & $0.098 \pm 0.004$ & $0.156 \pm 0.008$ \\   
               


 \cutinhead{\em Elliptical galaxies $R_{e}$ vs. $L_{deV}$} 

     All & 2861 & $0.739 \pm 0.013$ & $-6.992 \pm 0.140$ & $0.135 \pm 0.003$ & $0.168 \pm 0.004$ & $0.227 \pm 0.006$ \\   
     Sparse & 756 & $0.695 \pm 0.026$ & $-6.538 \pm 0.265$ & $0.148 \pm 0.006$ & $0.180 \pm 0.008$ & $0.259 \pm 0.011$ \\   
     Rich & 1313 & $0.815 \pm 0.023$ & $-7.792 \pm 0.239$ & $0.123 \pm 0.005$ & $0.158 \pm 0.006$ & $0.194 \pm 0.011$ \\


%

\enddata
\end{deluxetable}

\end{document}